\documentstyle[11pt,hrd_pasp,psfig,epsf,twoside]{article}
\newcommand{\itip}{$I^{TRGB}$}
\newcommand{\ome}{$\omega$ Centauri~}
\newcommand{\omen}{$\omega$ Centauri}
\newcommand{\mtip}{$M_I^{TRGB}$}

\def\edcomment#1{\iffalse\marginpar{\raggedright\sl#1\/}\else\relax\fi}
\reversemarginpar

\begin{document}
\title{A new calibration of the RGB Tip as a Standard Candle}
 \author{Michele Bellazzini, Francesco R. Ferraro}
\affil{Osservatorio Astronomico di Bologna, Via Ranzani 1, 40127, Bologna, 
Italy}
\author{Elena Pancino\altaffilmark{1}}
\affil{European Southern Observatory, K. Schwarzschild Str. 2, 
Garching, D-85748, Germany}

\altaffiltext{1}{on leave from Dip. di Astronomia,
Universit\`a di Bologna, Via Ranzani 1, 40127, Bologna, ITALY}

\begin{abstract}
We have obtained an accurate estimate of the absolute I magnitude of the Tip of
the Red Giant Branch (\mtip) for
the globular cluster \omen, based {\em (a)} on the largest photometric
database ever assembled for a globular, by Pancino et al. (2000), and {\em (b)}
on a direct distance estimate for \omen, recently obtained by
Thompson et al.(2001) from a detached eclipsing binary. 
The derived value \mtip
$=-4.04\pm 0.12$ provides, at present, the most accurate empirical
zero-point for the calibration of the \mtip -- [Fe/H] relation, at
[Fe/H]$\sim -1.7$.\end{abstract}

\section{Introduction}

While the use of the luminosity of the Tip of the Red Giant Branch
(TRGB) as a standard candle dates back to 1930 (see Madore \& Freedman 1998 
and references therein), the development of the method as a safe and
viable technique is relatively recent (Lee, Freedman \& Madore 1993, hereafter
L93). In a few years it has become a widely adopted technique, finding
fruitful application also within the {\em HST Key project on the
Extragalactic Distance Scale}. 
The underlying physical processes are clearly
understood and well rooted in the theory of stellar evolution
(Madore \& Freedman 1998). 
The method is particularly useful to estimate distances
to those stellar systems that do not contain Cepheids, such as early
type galaxies, and it can be applied to galaxies as far as $\sim 12$
Mpc with the current instrumentation. The key observable quantity is
the magnitude of the bright end (the tip) of the Red Giant Branch
(RGB), that corresponds to a sharp cut-off in the RGB Luminosity
Function (LF), measured in the Cousin's $I$
passband.In this passband the magnitude of the tip shows a very
weak (if any) dependence on metallicity (Da Costa \& Armandroff 1990, hereafter
DA90). The feature can be identified by applying the Sobel's filter,
an edge-detection algorithm, to the LF of the upper RGB (see
Sakai, Madore \& Freedman 1996, hereafter S96, for a standard application). 
The possible biases have been well
characterized and quantified by means of numerical simulations by
Madore \& Freedman (1995; hereafter MF95).

\begin{figure*}[t]
\plotfiddle{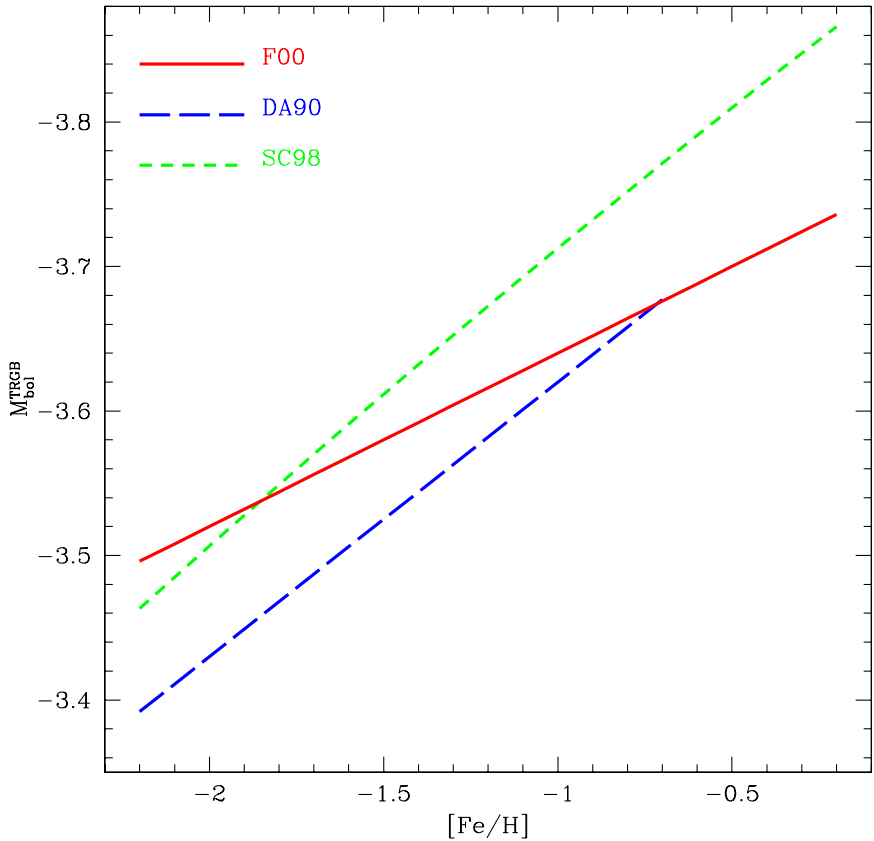}{11 truecm}{0}{60}{60}{-198}{-90}
\caption{Comparison between different relations between $M_{bol}^{TRGB}$ and
[Fe/H], providing the basis for the calibration of the \mtip vs. [Fe/H]
relations. The empirical calibrations by DA90 and F00 are compared to the
theoretical calibration by SC98.}
\end{figure*}

In this context, we are performing an extensive study of the Red Giant
population in Galactic Globular Clusters (GGC). In particular, we have
derived an accurate calibration of the RGB photometric properties as a
function of metallicity, both in the optical and in the IR
(Ferraro et al. 1999, 2000). The final aim of this program is to use stellar
populations in GGCs as {\it calibrators} for the TRGB method, taking
advantage of the large samples that can be easily assembled with the
new generation of array detectors and cameras.  As a part of this
general project, we present {\em (i)} a very robust
empirical calibration of \mtip at $[Fe/H]\sim -1.7$, derived by
applying the {\em standard analysis} to a very large sample of RGB
stars in the globular cluster \ome (NGC~5139), and based on a direct
distance estimate recently obtained by Thompson et al. (2001, hereafter T01)
for a detached eclipsing binary in this cluster, and {\em (ii)} a new
{\em empirical} calibration of the \mtip -- $[Fe/H]$ relation based on
a large, homogeneous IR database of RGB stars in Galactic Globular
Clusters, recently published by Ferraro et al. (2000; hereafter F00).

The results of this analysis are reported in full detail in Bellazzini, Ferraro
\& Pancino (2001; hereafter BFP01).

\section{The \mtip vs [Fe/H] relation}

The most widely adopted calibration of the absolute I magnitude of the TRGB is
based on small samples of RGB stars observed in a few templates globulars
(in the range $-2.2\le [Fe/H]\le -0.7$), and
relies on the RR Lyrae distance scale (L93, DA90). Such calibration suffer from
two major sources of uncertainty:

\begin{itemize}

\item{} The evolution of stars along the RGB becomes faster as their
luminosity increases along the path toward the TRGB. Thus, most of the
cluster light has to be observed to correctly sample the fastest
evolutionary phases. MF95 stated that acceptable
detections of the TRGB can be obtained if more than 50 stars are
sampled within 1 mag from the tip and that fine estimates can be
obtained only sampling more than 100 stars in that range. The DA90
samples are much poorer than this, therefore their \itip estimates may
be affected by the systematics associated with small number
statistics.

\item{} The RR Lyrae distance scale is quite uncertain. While there is now some
agreement on the slope of the $M_V(RRLy)$ -- $[Fe/H]$ relation, the
actual zero points is still hardly debated (see Cacciari 1999, for a
recent review).
   
\end{itemize}

Ferraro et al (2000; hereafter F00) have recently presented a similar 
calibration based on NIR photometry of
nine templates globular clusters. With respect to LF93 the calibration of F00 
(1) is based on larger samples of RGB stars, (2) it is less affected by
uncertainties in the reddening, because of the adopted NIR passbands, and (3) it
extends the range of application to $-2.2\le [Fe/H]\le -0.2$. The final
calibrating relation for \mtip is (see BFP01):

\begin{figure*}[t]
\plotfiddle{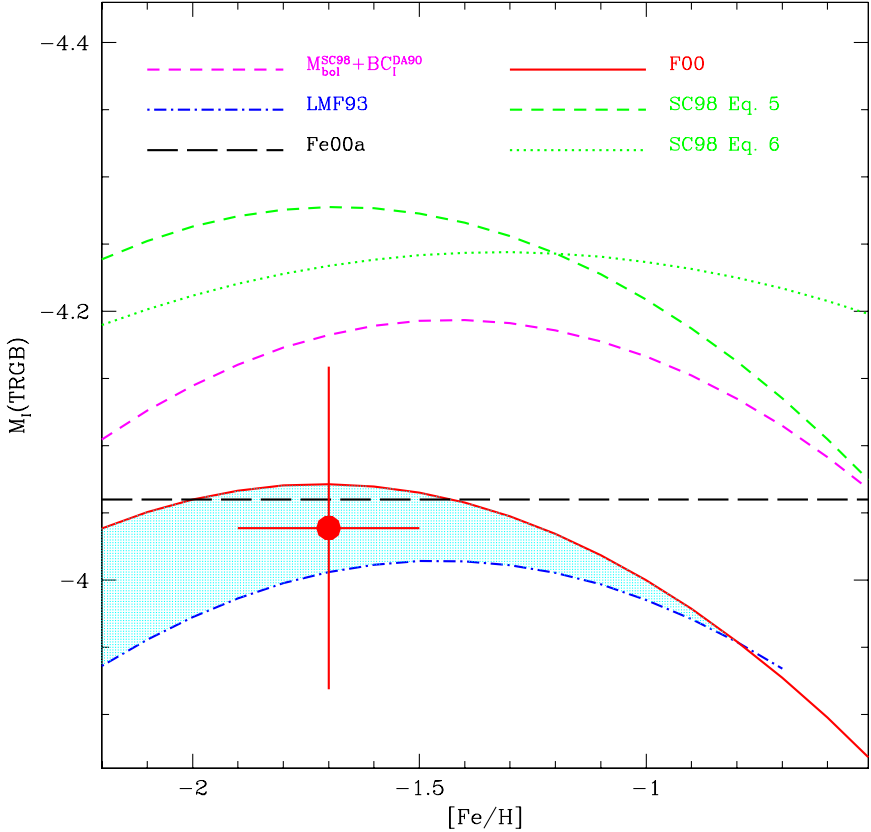}{11 truecm}{0}{60}{60}{-198}{-90}
\caption{Comparison between the calibration of \mtip\ obtained for \ome and
different \mtip\ -- [Fe/H] relations.   
The short dashed curve between
the SC98 relations and the shaded area (see text) is the calibration obtained 
by coupling the  $M_{bol}^{TRGB}$ vs. [Fe/H] relation by SC98 and the 
empirical $BC_I$ by DA90.
}
\end{figure*}

\begin{equation}
M_{I}^{TRGB}=0.14[Fe/H]^{2}+0.48[Fe/H]-3.66 
\end{equation}
 
Eq. (1) represents a substantial improvement with respect to
previous work, since it is based on the largest RGB samples in GGCs
available in the literature.  Note that the F00 survey sampled a
significant fraction of the cluster light (up to $\sim 80\%$).
However even in this large data-set the number of RGB stars in the
upper 1 mag bin from the TRGB is $< 50$ in all the cases, and the
brightest star detection is possibly prone to low number statistics
effects.  This is due, in most cases, to intrinsic poorness of the
cluster population, since the majority of GGCs simply {\em do not
contain a sufficient number of RGB stars} to adequately sample the
upper RGB. Thus extensive observations of the most massive GGCs are
urged in order to properly calibrate the above relation. 

\section{A firm zero point for the TRGB calibration}

\ome is a nearby and well studied cluster. It is the most luminous  
globular cluster in the Milky Way system, therefore even the fastest
evolutionary phases are well populated and it is possible to observe a
quite large sample of RGB stars, fulfilling the prescriptions of
MF95. Here we adopt the huge photometric ($B$,$I$) database presented
by P00, consisting of more than 220,000 stars observed with the WFI
camera at the 2.2 ESO-MPG telescope and extending over a $\sim
34^{\prime} \times 33^{\prime}$ field roughly centered on the cluster.
The absolute photometric calibration is accurate to within $\pm 0.02$
mag.  The observed field extends over $\sim 24 $ core radii, 
i.e. an area enclosing $90 \%$ of the cluster light.
Thus, virtually {\em all the bright stars of \ome are included in the
adopted database.} Considering that \ome is the most luminous globular
cluster of the whole Galaxy, this photometric database is the largest
sample that can be obtained in the Galactic globular cluster system.
 
Moreover, \ome was the first globular in which a detached eclipsing
binary system, {\it OGLEGC17}, was discovered (Kaluzny et al. 1996).  This
enabled T01 to obtain a direct distance estimate, independent of any
other distance scale. This method is basically {\em
geometrical}, since the distance is obtained by comparison between the
linear and angular size of the binary members (see
Kruszewski and Semeniuk 1999, for a recent review and references). 
The final distance obtained by T01 is $d= 5385\pm
300$ pc, which corresponds to $(m-M)_0=13.65\pm 0.11$, if we assume
their adopted extinction value, $E(B-V)=0.13\pm 0.02$. Their result is
in good agreement with previous estimates. According to T01, the error
bar on the distance modulus can be significantly reduced as soon as
better light and velocity curves will be obtained for {\it
OGLEGC17}. Thus, a significant improvement of the quoted measure of
the distance modulus has to be expected in the near future.  

The huge P00 sample allowed us to succesfully perform {\em the first measure of
the TRGB in a Globular Cluster using the Standard Technique}, i.e. by detecting
the cut-off of the RGB luminosity function with an edge-detector filter (see
BFP01 and references therein). Coupling our optimal measure of the TRGB with the
distance and reddening presented by T01 we obtained:

\begin{equation}
M_{I}^{TRGB}= -4.04 \pm 0.12
\end{equation}

where {\em all} the sources of uncertainty have been taken into
account. The main contributor to the error budget remains the estimate
of the distance modulus. Even at
the present level of accuracy, the \mtip measure derived here is the
less uncertain calibrating point for the \mtip -- [Fe/H] relation.

\section{Comparisons with other calibrations}

In Figure 2 our estimate of \mtip\ in \ome is reported as a {\em big
black dot} in the \mtip vs. [Fe/H] plane. The horizontal error bar
represents the range in metallicity covered by the dominant population
of the cluster (see BFP01).

The {\em heavy solid curve} is the new calibration based on the large
IR database by F00, while the L93 calibration is plotted as a
{\em dotted-dashed curve}. The horizontal {\em long dashed line}
represents the recent result by Ferrarese et al. (2000 hereafter Fe00a) who
calibrated the TRGB as a secondary indicator by using Cepheid
distances in a small set of nearby galaxies where both Cepheids and
the TRGB have been detected. They found \mtip $=-4.06\pm 0.07$
(random) $\pm 0.13$ (systematic), in very good agreement with our
estimate, despite their larger uncertainty. We plotted also
the region delimited by the two empirical calibrations 
as a {\it shaded area}, in order to point out the region of the plane
where most of the empirical estimates lie.

Salaris \& Cassisi (1998; hereafter SC98) provided a theoretical relation 
of \mtip\ as a function of metallicity (see BFP01). 
Using two different bolometric
corrections, SC98 derived two slightly different relations (their
eq. 5 and 6), which are reported in Fig. 2 as a {\em short dashed
curve} and as a {\em dotted curve}, respectively.

As can be seen from Fig. 2, the {\em theoretical} calibrations are
systematically ($\sim 0.2$ mag) brighter than {\em empirical} ones, as
already noted by SC98 and Fe00a. The strong constraint provided by the
TRGB luminosity in \ome seems to favor the empirical calibrations.

It is interesting to note that if the theoretical $M_{bol}^{TRGB}$ vs. [Fe/H]
relation by SC98 is coupled with the empirical $BC_I$ by DA90 the difference 
with
the purely empirical calibrations (shaded area) is reduced. This fact suggests
that the uncertainties in the bolometric corrections may significantly
contribute to the observed mismatch between the theoretical predictions and the
\ome observed point, which is {\em fully independent} from $BC_I$.

\acknowledgments
This research has been supported by the Italian Ministero della
Universit\`a e della Ricerca Scientifica e Tecnologica (MURST),
through the COFIN grant p. MM02241491\_004, assigned to the project
{\em Stellar Observables of Cosmological Relevance}.  The financial
support of the Agenzia Spaziale Italiana (ASI) is also kindly
acknowledged.

\end{document}